\documentclass[12pt]{article}

\usepackage{epsfig, amssymb, amsmath}
\DeclareGraphicsExtensions{.jpg, .eps}
\newcommand{\be}{\begin{equation}}
\newcommand{\ee}{\end{equation}}
\newcommand{\bea}{\begin{eqnarray}}
\newcommand{\eea}{\end{eqnarray}}
\newcommand{\ba}{\begin{array}}
\newcommand{\ea}{\end{array}}

\def\bbox{{\,\lower0.9pt\vbox{\hrule \hbox{\vrule height 0.2 cm
\hskip 0.2 cm \vrule height 0.2 cm}\hrule}\,}}
\newcommand{\dsl}{\pa \kern-0.5em /}

\newcommand{\tdu}{\tilde U}
\newcommand{\tdv}{\tilde V}
\newcommand{\U }{{\cal U} }
\newcommand{\V }{{\cal V} }

\font\mybb=msbm10 at 12pt
\def\bb#1{\hbox{\mybb#1}}

\def\bR {\bb{R}}

\def\appendix#1{
  \addtocounter{section}{1}
  \setcounter{equation}{0}
  \renewcommand{\thesection}{\Alph{section}}
  \section*{Appendix \thesection\protect\indent \parbox[t]{11.15cm}
  {#1} }
  \addcontentsline{toc}{section}{Appendix \thesection\ \ \ #1}
  }

\begin{document}

\begin{flushright}
\small
UG-05-05\\
UB-ECM-PF-05/19\\
DAMTP-2005-65\\
KCL-MTH-05-10\\
{\bf hep-th/0507143}\\
\date \\
\normalsize
\end{flushright}

\begin{center}


\vspace{.7cm}

{\Large {\bf Classical resolution of  singularities\\ 

\smallskip

in dilaton cosmologies}}

\bigskip

\vspace{1.2cm}

{\large E.A.~Bergshoeff${}^*$, A.~Collinucci${}^*$,
D.~Roest${}^\dagger$,}

{\large J.G. Russo${}^\ddagger$
and P.K.~Townsend${}^{\diamond}$
} \vskip 1truecm

\small
${}^*$\,{Centre for Theoretical Physics, University of
Groningen,\\
   Nijenborgh 4, 9747 AG Groningen, The Netherlands}\\
\vskip .3truecm

${}^\dagger$\,{Department of Mathematics, King's College,\\
London, Strand, London WC2R 2LS}
\vskip .3truecm

${}^\ddagger$\,{Instituci\'o Catalana de Recerca i
Estudis Avan\c{c}ats,\\
Departament ECM, Facultat de F\'isica,\\
Universitat de Barcelona, Spain}
\vskip .3truecm

${}^\diamond$\,{Department of Applied Mathematics and
Theoretical Physics\\
Centre for Mathematical Sciences, University of Cambridge\\
Wilberforce Road, Cambridge, CB3 0WA, UK}
\vspace{.7cm}


{\bf Abstract}

\end{center}

\begin{quotation}

\small

For models of dilaton-gravity with a possible exponential potential, such as the tensor-scalar 
sector of IIA supergravity,  we show how cosmological solutions correspond to trajectories in a
2D Milne space (parametrized by the dilaton and the scale factor). Cosmological singularities correspond to points at which a trajectory meets the Milne horizon, but the trajectories can be 
smoothly continued through the horizon to an instanton solution of the Euclidean theory. 
We find some exact cosmology/instanton solutions that lift to black holes in one higher 
dimension. For one such solution,  the singularities of a big crunch to big bang transition mediated 
by an instanton phase lift to the black hole and cosmological horizons of de Sitter 
Schwarzschild spacetimes.

\end{quotation}

\newpage

\section{Introduction}
\setcounter{equation}{0}

In models of gravity coupled to scalar fields, homogeneous and isotropic cosmologies correspond to trajectories  in an  `augmented target space'  or `superspace',  parametrized  by the scalar fields and the FLRW scale factor. There is a natural conformal class of Lorentzian  metrics on this space, the scale factor playing the role of `time', and when the scalar field target space is a hyperboloid this Lorentzian metric can be chosen to be the metric of a Milne `universe', which is just Minkowski space in polar-type coordinates. In this case, cosmological singularities correspond to points at which the trajectory reaches the Milne horizon \cite{Russo:2004am}.  For the models considered in \cite{Russo:2004am},  
the trajectories were actually geodesics and hence straight lines in Minkowski space that typically cross the Milne horizon twice, corresponding to a big bang and a big crunch singularity. In particular, there are trajectories that represent a universe undergoing a transition from a collapsing big-crunch universe to an expanding big-bang universe across a `forbidden' region of `superspace' behind the Milne horizon. 

In the absence of a potential for the scalar fields, it turns out that FLRW cosmologies are geodesics in the Milne metric only for a particular radius of curvature of the target space hyperboloid. We may parametrize these models by a constant $\gamma$ such that the motion is geodesic for $\gamma=1$. In a previous article we generalized the  construction of  \cite{Russo:2004am} to 
(generically) non-geodesic Milne motion in models for which the target space hyperboloid has an arbitrary finite radius, corresponding to arbitrary non-zero $\gamma$, and we showed  that parts of a trajectory behind the Milne horizon may be interpreted as instanton solutions of the Euclidean theory \cite{Bergshoeff:2005cp}. We also showed that for $\gamma=2$ models there exist  trajectories that are smooth closed curves in the Minkowski extension of the Milne superspace, corresponding to cyclic universes in which cosmological `phases'  are separated by  instanton `phases',  each of which mediates a big crunch to big bang transition across a region behind the Milne horizon\footnote{Cyclic  universes without cosmological singularities are possible in some models  with both a hyperbolic target space and a scalar potential \cite{Billyard:2000cz} but these have no instanton `phases' and are therefore quite different from the cyclic universes under discussion here.}. 

These ideas are applicable to IIB supergravity because the dilaton and axion of that theory parametrize the hyperbolic space $Sl(2;\bR)/SO(2)\equiv H_2$, and they potentially imply a resolution
of singularities of flat ($k=0$) cosmologies in IIB superstring theory since these correspond to null lines in the Minkowski extension of the Milne superspace. However, the $H_2$ radius of curvature (which corresponds to $\gamma=2/3$) is not in the range for which a generic trajectory has a smooth continuation through the Milne horizon for $k\ne0$, and the same applies to the models in spacetime dimension $d<10$ obtained by toroidal compactification.

One purpose of this paper is to apply these ideas  to models of dilaton-gravity such as the dilaton-gravity sector of IIA supergravity. As already pointed out in \cite{Bergshoeff:2005cp}, cosmological solutions of  IIB supergravity for which the axion is
identically zero are also solutions of IIA supergravity. 
More generally, one could 
set the axion to zero after having redefined the fields by an
$Sl(2;\bR)$ transformation. This means that
 any {\it planar} cosmological trajectory in the 3D Minkowski augmented  target space of IIB supergravity must correspond
to some solution of the scalar-tensor sector  of IIA supergravity, so this already shows that the latter
can be viewed as trajectories in a 2D Minkowski space. 
The possible consistent truncations simply
correspond to the possible choices of this 2D Minkowski subspace.
A subtlety that arises in these truncations
is that the parameter $\gamma$ ceases to be physical because the truncated target space is now a  
``one-dimensional hyperboloid''  for which the intrinsic curvature is trivially zero. Thus, planar 
trajectories for various values of $\gamma$ that previously corresponded to different solutions of 
different models must now be viewed as different solutions of the {\it same} model. This leads to  ambiguities in the continuation of trajectories across Milne horizons. We investigate the consequences of this in some detail.

A further feature of IIA supergravity is that it allows an extension to a ``massive'' IIA theory with an exponential dilaton potential. This motivates the investigation of cosmological trajectories in dilaton-gravity models in $d$ spacetime dimensions with an exponential potential
characterized by an arbitrary dilaton coupling constant $a$.
Thus, our starting point will be the d-dimensional action
\be\label{laginitial}
I = \int d^d x\, \sqrt{\epsilon\det g}\left[ R - \tfrac{1}{2} (\partial \phi)^2
- \Lambda e^{-a\phi} \right]
\ee
for $d$-metric $g$ of signature $(\epsilon,1, \ldots , 1)$
and dilaton $\phi$, and constants $\Lambda$ and $a$.
We may suppose that $a\ge0$ without loss of generality, but allow $\Lambda$ to be positive, negative or zero. As in \cite{Bergshoeff:2005cp},
we have introduced a sign $\epsilon$ such that $\epsilon=1$ yields the
Euclidean Lagrangian\footnote{To have a  Euclidean action that is positive definite when $R=0$ one would need an additional overall factor of
 $-\epsilon $, but this does not affect the field equations, which
 is all that is needed here.}. For $\epsilon=-1$ and $d=10$ we have a
consistent truncation of IIA supergravity when $\Lambda=0$, and
when $\Lambda > 0$, we have a consistent truncation of massive
IIA supergravity when $a= 5/2$. 
Other values of $a$, for positive
and negative $\Lambda$, arise in M-theory compactifications to $d<10$.

These models have been extensively studied in the past, at least for $\epsilon=-1$. 
The equations describing cosmological solutions are known to be equivalent to those 
of an autonomous dynamical  system \cite{Halliwell:1986ja},  such that cosmologies correspond to trajectories in a 2D phase  plane\footnote{The phase plane should not be confused with the 2D Minkowski superspace; it is  essentially the space of velocities in this space for a particular time parametrization.}. This allows a determination of the  qualitative nature of the full space of 
solutions, and in this sense the cosmological solutions are already well understood. In particular, all $k=0$ cosmologies can be found exactly \cite{Burd:1988ss,Townsend:2003qv,Russo:2004ym}.
Here we provide a new `explanation'  for this fact: these solutions are  geodesics in the Milne wedge  of a 2D Minkowski space. This description allows  us to explore their continuation
through cosmological spacetime singularities as smooth trajectories through
the Milne horizon. 

A subset of the models studied here have the property that their action is the
reduction of the $(d+1)$-dimensional  Einstein-Hilbert action, with a possible
cosmological constant. This allows the $d$-dimensional cosmological solutions 
to be lifted to $(d+1)$-dimensional Einstein metrics. This is true not only for the cosmological
solutions but also for the instanton solutions because the $d$-dimensional 
Euclidean action is obtained in these cases by considering time-independent
fields. Thus, both cosmological and instanton solutions in $d$ dimensions lift to Lorentzian 
signature $(d+1)$-metrics. For example, when $\Lambda=0$ the action (\ref{laginitial}) is the 
dimensional reduction of the $(d+1)$ dimensional Einstein-Hilbert action.
As already mentioned in \cite{Bergshoeff:2005cp} it is known that $d$-dimensional cosmologies 
lift to the interior of a $(d+1)$-dimensional Schwarzschild black hole, and that 
instanton solutions lift to the Schwarzschild exterior. These results are applicable to IIA 
supergravity and they show that  cosmology or instanton solutions can be lifted to solutions of 
11D supergravity.  

Our interest here is the lift of  the 
full cosmology/instanton solutions corresponding to complete smooth 2D Minkowski trajectories. 
We find, in particular, that the $k=1$ Milne geodesics, 
 lift to a sequence of all the possible  $(d+1)$-dimensional (anti) de Sitter-Schwarzschild  spacetimes for a given absolute value  of the black hole mass, the negative and positive mass black holes corresponding to pre and post big-bang phases, respectively. Moreover, the big-crunch singularity is resolved in the
 higher dimension in the same way that certain dilaton black hole 
singularities are resolved \cite{Gibbons:1994vm}. 

When $\Lambda$ is non-zero there is a similar story for a particular ($d$-dependent) value of $a$ but
the $(d+1)$-dimensional action now has a cosmological constant. Exact cosmological and 
instanton solutions of the $d$-dimensional theory can be found for this value of $a$. It turns out that 
the $k=1$ cosmology/instanton solutions corresponding to 2D Minkowski trajectories that are smooth 
for $\gamma=1$ lift  to $(d+1)$-dimensional (anti) de
Sitter-Schwarzschild  spacetimes. The $\Lambda>0$ case is  of particular interest because there are then trajectories on which an entire instanton mediated big-bang to big-crunch transition becomes non-singular in the higher dimension, in the sense that the big-bang and 
big-crunch curvature singularities become regular horizons of the $(d+1)$-metric.

\section{Cosmologies and Instantons}
\label{sec.exact}
\setcounter{equation}{0}

We begin by  explaining how the problem of finding cosmological or instanton
solutions of the general model with action (\ref{laginitial}) can be reduced to solving the
equations of motion of an effective action describing the motion of a particle in a
2D Minkowski space. This is achieved by the ansatz
\be\label{ansatz}
ds^2_d = \epsilon \left(e^{\alpha\varphi}f\right)^2 d\tau^2 +
e^{2\alpha\varphi/(d-1)}d\Sigma_k^2 \, ,\qquad
\phi=\phi(\tau)\, ,
\ee
where $f$ is an arbitrary function of $\tau$, and
\be
\alpha = \sqrt{d-1\over 2(d-2)}\, .
\ee
The $(d-1)$-metric $d\Sigma_k^2$ is (at least locally) a maximally
symmetric space of positive ($k=1$), negative ($k=-1$) or
zero ($k=0$) curvature. One can choose coordinates
such that
\be
 d\Sigma_k^2 = (1-kr^2)^{-1}dr^2 + r^2 d\Omega_{d-2}^2\ ,
\label{euss}
\ee
where $d\Omega_{d-2}^2$ is an $SO(d-1)$-invariant metric on the
unit $(d-2)$-sphere; note that $d\Sigma_k^2 =d\Omega_{d-1}^2$
when $k=1$.
This ansatz constitutes a consistent
reduction of the original degrees of freedom to a two-dimensional
subspace, the `augmented target space', with coordinates
$(\varphi,\phi)$. The full equations of motion reduce to a set of
equations that can themselves be derived by variation (with respect to $\phi$, $\varphi$ and $f$) of the effective action
\be\label{originalact}
I= \int d\tau\,\left\{
 \tfrac{1}{2} f^{-1} \epsilon \left( \dot\varphi^2
- \dot\phi^2 \right) + f \left[
k(d-1)(d-2) e^{\varphi/\alpha} -
\Lambda e^{2\alpha\varphi - a\phi}\right] \right\}\,,
\ee
where the overdot indicates differentiation with respect to $\tau$.
For $\epsilon=-1$ we can interpret  $\tau$
as a time coordinate related to the
time $t$ of FLRW cosmology in standard coordinates by
\be\label{FLRWtime}
dt = e^{\alpha\varphi} f d\tau\, .
\ee
For $\epsilon =1$ the metric has Euclidean
signature and we can interpret
$\tau$ as imaginary time.

To proceed we introduce new field variables $(U,V)$ that we will shortly be able to interpret as
null coordinates for a 2D Minkowski space. These variables are defined by
 \be
 e^\varphi = (-\epsilon U V)^{\alpha \gamma} \,, \qquad e^\phi = \left(- \frac{\epsilon V}{U} \right)^{\alpha \gamma} \,,
\label{kop} 
\ee
for some constant $\gamma$. Although different choices of $\gamma $
give equivalent solutions in a given patch, say $U>0,\ V>0$, $\epsilon=-1$,
the continuation of these solutions to the entire Minkowski plane $(U,V)$
is sensitive to the value of $\gamma $,
as will be explained in section 3. 
Note that we need both signs of $\epsilon$ to cover the entire Minkowski plane $(U,V)$ with real scalars
$(\varphi, \phi)$: if the coordinates  $U$ and $V$ have the same sign we have to choose
$\epsilon = -1$ and if they have opposite signs $\epsilon = 1$. 
With this proviso one can always assume both
$(-\epsilon U V)$ and $(-\epsilon V / U)$ to be positive.
In terms of the new variables, the effective action is
\bea
I &=& \int d\tau\,\Bigg\{ 2 \alpha^2 \gamma^2 \epsilon
(f\, UV)^{-1} \dot U \dot V 
\nonumber\\
&&+ \
f\left[k (d-1)(d-2) (- \epsilon UV)^{\gamma}
-  \Lambda (-\epsilon UV)^{2 \alpha^2 \gamma} (-\epsilon U /V)^{\alpha \gamma a}
\right] \Bigg\}\,.
\eea

To obtain a canonical kinetic term we now fix the time reparametrization invariance  by choosing
\be
 f = -{2 \over (d-2) \epsilon UV}\, .
\ee
The resulting action is
 \bea\label{masteract}
  I &=& -{1\over2}(d-1)\int d\tau\,\Big\{ \gamma^2 \dot U \dot V -  4 k (-\epsilon U V)^{\gamma - 1}\nonumber\\
 &&\qquad\qquad +\  \frac{4 \Lambda}{(d-1)(d-2)} (-\epsilon U V)^{2 \alpha^2 \gamma  -1}
 (-\epsilon U / V)^{\alpha \gamma a}  \Big\}\, .
\eea
However, the equations of motion of this action must be supplemented
by the $f$ equation of motion of the original action, which is the constraint
\be
   \gamma^2 \dot U \dot V = -4 k (-\epsilon U V)^{\gamma - 1}
  + \frac{4 \Lambda}{(d-1)(d-2)} (-\epsilon U V)^{2 \alpha^2 \gamma  -1}
 (-\epsilon U / V)^{\alpha \gamma a} \,.
 \ee
Note that this is just the condition that a particle in a 2D space
with action (\ref{masteract}) have zero energy. Given a zero-energy
solution of the equations of motion for $U$ and $V$ we may read off the
(Einstein-frame) metric and dilaton field from
\bea\label{einstein}
  ds^2_d &=& \frac{4 \epsilon}{(d-2)^2} (-\epsilon U V)^{2 \alpha^2
    \gamma - 2} d\tau^2 + (-\epsilon U V)^{2 \alpha^2 \gamma / (d-1)}
  d\Sigma_k^2\ ,
 \nonumber\\
  e^\phi &=& \left( -\frac{\epsilon V}{U} \right)^{\alpha \gamma}  \,.
\eea
For $\epsilon=-1$, the FLRW time $t$ is now related to the
parameter $\tau$ by
 \be \label{FLRWtime2}
  dt  \propto ( U V)^{\alpha^2 \gamma - 1} d \tau \,.
 \ee

We have now formulated the problem in such a way that
cosmologies, given by (\ref{einstein}),  correspond to trajectories
in a 2D Minkowski space with null coordinates $U,V$. 
More precisely, cosmological solutions correspond to trajectories in either
the future Milne patch ($U>0,\, V>0$) or the past Milne patch ($U<0,\,
V<0$) of the 2D Minkowski space. The null lines $U=0$ and $V=0$
are the Milne horizon. The 2D Minkowski space should therefore
be thought of as the analytic extension of a Milne ``universe''.
Trajectories in this ``universe'' correspond to cosmologies, and
a cosmological singularity in spacetime corresponds to a point at
which a trajectory crosses the Milne horizon. On crossing the horizon, into the 
Rindler patches of the 2D Minkowski space, the trajectory must be 
re-interpreted as an `instanton', i.e. as a solution of the Euclidean action. 

Despite the cosmological spacetime singularity, the continuation of the 
cosmological trajectory through the Milne horizon may be smooth. 
For example,  if $a=0$ and $k=0$ we may choose $\gamma$
so as to make the potential a constant, in which case
the motion is geodesic motion and all trajectories are straight
lines in the 2D Minkowski `superspace'. This is essentially the point made in
\cite{Russo:2004am} in the context of a model with $N$ scalar
fields, in which case the trajectories were straight lines in an
$(N+1)$-dimensional Minkowski space. Here we are restricted to $N=2$ but
we allow for non-zero $a$ and consider non-flat ($k\ne0$)
cosmologies. Clearly, in this more general context the motion in the
2D Minkowski space will not be geodesic but there may nevertheless
be smooth trajectories that cross the Milne horizon. From
\cite{Bergshoeff:2005cp} we know that this
indeed happens when $\Lambda=0$ and even that there can be elliptical
trajectories in the analytic continuation of Milne to Minkowski
space that correspond to a closed ($k=1$) universe undergoing
an endless cycle of big-bang to big crunch transitions mediated
by ($k=-1$) instanton ``phases''.

We now plan to investigate to what extent these features carry over
to the general model summarized by the effective action
(\ref{masteract}). We will concentrate on those special cases for
which an exact solution can be found. These cases are as follows:

\begin{itemize}
\item  $\Lambda=0$. This case was partly dealt with in
\cite{Bergshoeff:2005cp}, as the special case of planar trajectories
in a IIB-type model. In the IIA-type case considered here, the
Lagrangian  (\ref{laginitial}) from which we started is the
dimensional reduction of the Einstein-Hilbert action in $(d+1)$
dimensions. The dimensional-reduction formula is
 \bea
  ds^2_{d+1} &=& - \epsilon e^{\phi / \alpha} dz^2 + e^{-2 \alpha \phi / (d-1)} ds_d{}^2  \nonumber \\
  &=& - \epsilon \left( -\frac{\epsilon V}{U} \right)^{\gamma} dz^2
      + \frac{4 \epsilon}{(d-2)^2} (-\epsilon UV)^{2 \alpha^2 \gamma -2} \left( -\frac{\epsilon U}{V} \right)^{2 \alpha^2 \gamma / (d-1)} d \tau^2
       \nonumber \\
  && +\ (U^2)^{2 \alpha^2 \gamma / (d-1)} d\Sigma_k{}^2 \,.
 \label{xlift}
 \eea
Note that this metric has Lorentzian signature for either sign of
$\epsilon $. 
This allows any cosmology/instanton solution of the $d$-dimensional theory 
to be re-interpreted as a  Lorentzian signature solution of $(d+1)$-dimensional 
General Relativity.

\item  $a=\sqrt{2}/\sqrt{(d-2)(d-1)}$. For this value of $a$ the lagrangian
(\ref{laginitial}) is the dimensional reduction of $d+1$-dimensional
gravity with a cosmological constant, so we can use the $d+1$ dimensional
Schwarzschild (anti) de Sitter solution to obtain a solution in $d$ dimensions.

\item  $a=0$. In this case, the motion in 2D Minkowski space is analogous 
to motion in a central potential.

\item  $k=0$. In this case all cosmological solutions are already known
explicitly. Here we give an alternative description of them as
straight-line geodesics in a 2D Milne `superspace'.

\end{itemize}

\section{Zero cosmological constant ($\Lambda=0$)} \setcounter{equation}{0}

Setting $\Lambda=0$ in (\ref{masteract}) we get the action
 \be\label{zerolam}
  I =  -\tfrac{1}{2} (d-1)\int d\tau\,\Big\{ \gamma^2 \dot U \dot V
 -  4 k  (-\epsilon U V)^{\gamma - 1} \Big\}\, .
 \ee
We have still to choose $\gamma$. The choices $\gamma=1$ and
$\gamma=2$ are special because the equations of motion are then
linear, so we shall consider only these two possibilities. From
(\ref{kop}) it is obvious that
the variables $(U,V)$ for the choice $\gamma=1$ are not the same as
these  variables for the choice $\gamma=2$, so we rename
the latter $(\tdu ,\tdv )$. Also, it is clear from
(\ref{FLRWtime2}) that the independent parameter $\tau $ for the choice
$\gamma=1$ is not the same as the independent parameter for the choice
$\gamma=2$, so we call the latter $\lambda $.
These considerations lead us to consider the following cases:
\begin{itemize}

\item $\gamma =1$. In this case we take the
independent variable to be $\tau$ and
the dependent variables to be $( U,V)$, so we seek
solutions of the equations of motion
for $ U(\tau)$ and $V(\tau)$.

\item $\gamma =2$. In this case we take the independent variable to be $\lambda$ and
the dependent variables to be $(\tdu,\tdv )$. In other words,
we seek solutions of the equations of motion for $\tdu (\lambda)$ and
$\tdv (\lambda)$. Note that, the formula (\ref{einstein}) for the metric and dilaton 
still applies but with $(U,V)$ replaced by $(\tdu,\tdv)$ and $\tau$ replaced 
by $\lambda$.

\end{itemize}

For either choice, the cosmological solutions correspond to
trajectories in a Milne wedge of 2D Minkowski space with
$UV>0$, or $\tdu \tdv >0$, since this condition is required for
real $\varphi$. The null lines $UV=0$ (or $\tdu\tdv =0$) are
the Milne horizon, and cosmological singularities correspond to points
at which this horizon is crossed. This can be seen from the fact that
the scale factor $e^\varphi$ goes to zero as $UV\to 0$. Inasmuch as
we are concerned only with cosmological solutions away from their
big-bang and big-crunch singularies, the two choices of $\gamma$ yield
equivalent solutions; the correspondence between them is given by
\be
| U(\tau )| = \tdu ^2(\lambda )\ , \qquad |V(\tau )|=\tdv ^2(\lambda )\ ,
\label{ert}
\ee
and the relation
\be\label{tautolam}
d\tau = -\epsilon \tdu (\lambda)\tdv (\lambda) d\lambda\, .
\ee
We shall solve the equations of motion for both values of $\gamma$ and
verify that the solutions are related in this way.

Given this relation, it might seem pointless to consider both values of
$\gamma$. However, the two choices of $\gamma$ need not yield equivalent
trajectories in the analytic continuation of 2D Milne to 2D Minkowski
space. The reason for this is that the Minkowski space with null coordinates
$(U,V)$ is {\it not the same space} as the Minkowski space with null coordinates
$(\tdu ,\tdv )$; the two differ by a conformal transformation
that is singular on the Milne horizon. This possibility is a special
feature of 2D Minkowski space: the conformal transformation can
be removed by a time reparametrization, so that what was initially
motion in a 2D Minkowski space again becomes motion in a 2D Minkowski
space, albeit with a different potential energy function. Thus, there
is more than one possible continuation of a cosmological
trajectory to a smooth trajectory in the 2D Minkowski continuation
of 2D Milne. Moreover, a smooth trajectory that crosses the Milne
horizon in the coordinates $(U,V)$ will not be smooth in the
coordinates $(\tdu ,\tdv)$ because of the singularity of the
conformal rescaling that relates the Minkowski metrics in these two
sets of variables.

\subsection{Milne Geodesics ($\gamma=1$)}

We now set $\gamma = 1$ in (\ref{zerolam}) to get the action
 \be
  I = -\tfrac{1}{2} (d-1)\int d \tau \, \Big\{  \dot  { U}\dot  { V} -  4 k  \Big\}\, .
 \ee
The potential is now a constant, so the equations of motion are
\be
\ddot   U = 0\ ,\qquad \ddot  V = 0\, .
\ee
The solutions, subject to the zero-energy constraint
\be
\dot   U\, \dot   V = -4 k \, ,
\ee
are the straight lines
\be \label{straight-lines}
  U = c \tau\ ,\quad  V= -{4 k\over c} \tau  + m\, , 
\ee
for constants $c$ and $m$. 
Each of these straight lines corresponds to some cosmology/instanton solution of the
equations of motion of our original action (\ref{laginitial}). It will
 be a cosmology in a region of the $( U, V)$ plane with 
$ U V>0$ and an instanton in a region with $ U 
 V<0$. 
Each of these solutions
must descend from a Lorentzian-signature $(d+1)$-dimensional metric, and we now
wish to determine what these higher-dimensional metrics are. 

We may restrict ourselves here to $k\ne0$ as the general $k=0$ case will be discussed
later in a separate section, so this leaves $k=\pm1$. The $m =0$ case is clearly special, 
as the straight line trajectory passes through the origin of the $( U, V)$ space, and $k\epsilon=1$ everywhere;  in these cases the dilaton is constant and the $d$-metric is flat (for $k=-1$ it is the 
$d$-dimensional Milne metric). In addition, the parameter $c$ can be
adjusted by rescaling $ U\to c_0  U,\ 
 V\to   V/c_0$,  so we choose $c=2$.
One then finds that the $(d+1)$-metric
(\ref{xlift})
is
\be
ds^2_{d+1} = k dr^2 + r^2d\Sigma_k^2  - {k}dz^2\, , 
\ee
where
\be
r^{d-2} = 2\tau\, . 
\label{rrr}
\ee
For either sign of $k$, this is just a flat metric on $(d+1)$-dimensional Minkowski space.
Thus  the spacetime singularity at the Milne superspace horizon
lifts in this case to a mere  coordinate singularity at $r=0$.

Let us now consider the less trivial case of $m \ne 0$. We shall assume that 
$m >0$ and  $c>0$ since the other cases are entirely analogous. 
For $k=1$ the straight line trajectory originates at $\tau=-\infty $ in the Rindler patch $ U<0,\,
  V>0$, where we must choose $\epsilon=1$, so this part of the
 trajectory corresponds to an instanton solution.
When the straight line crosses the $ U=0$ horizon it enters the Milne
 patch, where we must choose $\epsilon =-1$, so this part of the
 trajectory corresponds to a cosmological solution.
Finally, the straight line passes through  the $ V=0$ horizon to
 enter the other Rindler patch ($ U>0,\,  V<0$) where we must again
 choose   $\epsilon=1$. So  this part of the trajectory describes another 
 instanton solution. This is illustrated in Fig.1a.
For $k=-1$ we again have three solutions corresponding to three pieces 
 of the straight line but now two are cosmologies and one an instanton; the instanton mediates
 between a collapsing big-crunch universe and an expanding big-bang universe.

 Thus a single straight-line trajectory comprises one cosmology and two instantons (for $k=1$) 
 or one instanton and two cosmologies (for $k=-1$),  all three of which lift, via the formula (\ref{xlift}),  
 to the Lorentzian-signature $(d+1)$-metric:
 \be  ds_{d+1}^2 = -\frac{4}{(d-2)^2}  { U \over  {V} }  
( U^2)
^{-(d-3)/(d-2)} d \tau^2 +  { V \over  {U} } dz^2  + ( U^2)^{1/(d-2)}
 d \Sigma_k{}^2 \, .
 \ee
To see how this metric incorporates all three of the $d$-dimensional 
cosmology/instanton solutions,  
we must consider how it looks in the various sectors of the 
 $ U, V$ plane through which a 
straight-line trajectory passes. For sectors with  $ {U}>0$
we may use  the coordinate $r$ defined in (\ref{rrr}), and we again set $c=2$,
in which case the $(d+1)$-metric becomes
 \be\label{Schw}
  ds_{d+1}^2 =  h(r)^{-1} d r^2 - h(r) dz^2 + r^2 d \Sigma_k{}^2 \,,
 \ee
with
 \be
  h(r)\equiv - { V \over  {U} }  = k  - \frac{m}{r^{d-2}} \,.
 \ee
 For sectors with $ U<0$, and again for $c=2$,  we may introduce a new coordinate  $\rho$ by setting 
 $2\tau=-\rho^{d-2}$, in which case the $(d+1)$-metric  becomes
 \be\label{negS}
   ds_{d+1}^2 =  g(\rho )^{-1} d {\rho}^2 - g(\rho) dz^2 + \rho^2 d \Sigma_k{}^2 \,,
 \ee
with 
 \be
  g(\rho ) = k + \frac{m}{\rho^{d-2}} \,.
 \ee

Now consider the $k=1$ case, for which $d \Sigma_k{}^2 =d\Omega_{d-1}^2$ (the 
 $SO(d)$-invariant metric on the unit $(d-1)$-sphere). In this case the metric (\ref{Schw}) is 
 just the Schwarzschild black hole in $d+1$ dimensions with mass
 proportional to $m$. 
This comprises the interior of the black hole ($ V>0$) and the
 exterior  ($ V<0$). 
Therefore 
the interior of the black hole describes the cosmology part of the
trajectory (similar to \cite{beh}) and the exterior
describes one of the two instanton parts of the trajectory. The other instanton part of the trajectory
lifts to the $k=1$ case of the metric (\ref{negS}), 
which is just Schwarzschild 
$(d+1)$-metric with a negative mass (proportional to $-m $). 
Thus, the three pieces of a single $k=1$ straight line trajectory in the 2D Minkowski `superspace' 
correspond to cosmological/instanton solutions that lift to the following three $(d+1)$-dimensional  Lorentzian-signature spacetimes:

\begin{itemize}

\item  negative mass Schwarzschild black hole.

\item  interior of positive mass Schwarzschild black hole.

\item exterior of positive mass Schwarzschild black hole.

\end{itemize}

The situation for $k=-1$ is analogous, but now
$d\Sigma_{k}=dH^2_{d-1}$ where $dH^2_{d-1}$ is the $SO(1,d-1)$
invariant metric on the unit radius $d-1$ hyperboloid.
 For example, the big-bang cosmology lifts to the $(d+1)$-metric 
\be
ds^2_{d+1}= - \left(1 + {m \over  r^{d-2}}\right)^{-1} dr^2 + 
 \left(1 + {m \over r^{d-2}}\right) dz^2+
r^2 dH^2_{d-1}  \, . 
 \ee
 Clearly, the coordinate $r$ is now a time coordinate, and the metric is asymptotic at late times 
 to the product of 
 a $d$-dimensional Milne universe with a circle. There is a big-bang singularity at $r=0$, which is a curvature singularity of the metric. Thus, the big-bang singularity of the $d$-dimensional metric is not resolved in the higher dimension (although the trajectory in `superspace' passes smoothly through it).  In contrast, the big crunch lifts  to the horizon of the $(d+1)$-metric
 \be
 ds^2_{d+1}= - \left(1 - {m \over  \rho^{d-2}}\right)^{-1} d\rho^2 
 +  \left(1 - {m \over  \rho ^{d-2}}\right) dz^2+ \rho^2dH^2_{d-1} \, .
\label{cvb}
\ee
The `exterior' spacetime ($\rho^{d-2}> m $) corresponds to the
$d$-dimensional big-crunch cosmology while the interior ($\rho ^{d-2}< m $) corresponds to the  $d$-dimensional instanton.

It should be appreciated that the asymmetry between the big-bang and
big-crunch singularities, for either sign of $k$, is due to an
asymmetry in the formula (\ref{xlift}) used to lift the
cosmology/instanton solutions to one higher dimension, and is not an
intrinsic feature of  the $d$-dimensional solutions. In fact, there are
two formulae that yield the same $d$-dimensional action; one is
related to the other by an interchange of $ U$ and $ V$ or, equivalently, 
by a flip of sign of the dilaton, which is a symmetry of the $\Lambda=0$ 
action. If the
other formula is used to lift to $d+1$ dimensions then one finds that
the black hole horizon and the curvature singularity behind the
horizon are exchanged, so either singularity can be resolved by an
appropriate lift\footnote{Note however, that the $\phi \to -\phi$ symmetry of the 
action that allows for this does not extend to the full IIA supergravity
action, nor to the case of non-zero $\Lambda$.}.

\subsection{Cyclic cosmologies ($\gamma=2$) }

We now set $\gamma =2$ in (\ref{zerolam}) to get the action
 \be
  I = - 2 (d-1) \int d\lambda\,\Big\{  \partial_\lambda  {\tdu } \partial_\lambda  { \tdv  } +  \epsilon k  \tdu  {\tdv } \Big\}\, ,
 \ee
where, according to the earlier discussion, we have replaced the
variables $(U,V)$ by $( \tdu , \tdv )$ and the independent
variable $\tau$ by $\lambda $.
The equations of motion are
 \be
 \partial_\lambda^2 {\tdu } = \epsilon k \, \tdu \, ,\qquad  \partial_\lambda^2 {\tdv  }= \epsilon k\,  \tdv \, ,
 \ee
and the zero-energy constraint is
\be
\partial_\lambda  {\tdu } \partial_\lambda  {\tdv } = \epsilon k \, \tdu \tdv \, .
\ee
For $\epsilon k=-1$ we get cyclic universes, as described  in
\cite{Bergshoeff:2005cp}.
The solutions are
\be\label{cyclic}
\tdu =\tdu _0\, \sin\lambda\ ,\qquad \tdv =\tdv _0\, \cos\lambda \ .
\ee
for constants $(\tdu _0,\tdv _0)$. A single cycle comprises four segments, joined smoothly across the Milne horizon at $\tdu \tdv =0$, one in each of the  following four segments of the 2D Minkowski analytic continuation of the 2D Milne wedge:
\bea\label{foursegments}
I: \ \tdu >0\, ,\ \tdv >0\, &;& \qquad II: \ \tdu >0\, ,\ \tdv <0 \nonumber\\
III:\ \tdu <0, \, \tdv <0\, &;& \qquad IV: \ \tdu <0\, , \tdv >0\, .
\eea

Using (\ref{ert}), and the relation (\ref{tautolam}),   we can make contact with the straight-line solutions of the previous subsection. Firstly, for the cyclic universe solution (\ref{cyclic}) the relation (\ref{tautolam}) can be integrated to give 
 \be
  \tau = -\tfrac{1}{2} \epsilon \tdu _0 \tdv _0 \sin^2 (\lambda) \,.
 \ee
Recalling that $k\epsilon=-1$ for the cyclic universe, we then deduce from (\ref{ert}) that the parameters $c,m$ of the corresponding $\gamma=1$ straight-line solutions (\ref{straight-lines})  are
\be
m= \mp \tdv_0^2\, , \qquad c = \mp {2\tdu_0\over \tdv_0}\, . 
\ee
This yields 
 \be\label{straightcyclic}
   U = \pm \epsilon \tdu _0{}^2 \sin^2 (\lambda) \, ,\qquad
   V = \mp \tdv _0{}^2  \cos^2 (\lambda) \,.
 \ee
Note that the slope of the straight line depends on $\epsilon$: 
upon crossing the Milne 
horizon there is a `kink' in the line. In this way, 
the elliptical trajectory in 
the $(\tdu ,\tdv )$-plane is mapped to a parallelogram with equal
sides, i.e. a rhombus,
in the $( U,  V)$-plane. The slope of the sides depends 
on the ratio $\tdu _0{}^2 / \tdv _0{}^2$, which determines the
eccentricity of the $\gamma=2 $ ellipse,
with a circle mapping to a square. 
This is a concrete illustration of how a trajectory  that is smooth for 
one choice of  $\gamma$ may have derivative discontinuities at the Milne horizon for 
some other value of $\gamma$.

Let us now investigate how the cyclic solutions look in $d+1$ dimensions.
Using the formula (\ref{xlift}), we get
\be
ds^2_{d+1}=\epsilon \left( 1- {\tdu _0^2\over r^{d-2}}\right) dy^2 -
\epsilon \left( 1- {\tdu _0^2\over r^{d-2}}\right)^{-1}dr^2
+r^2d\Sigma_k^2\ ,
\ee
where
\be
y={\tdv _0\over \tdu _0}\, z\ , \ \ \ \ r^{d-2}=\tdu _0^2\sin^2\lambda \ .
\ee
Each cosmology phase of the cycle has $\epsilon=-1$ and therefore
$k=1$, so it lifts to the interior of a 
Schwarzschild black hole, as in the previous subsection.
These are the segments of the ellipse lying on the Milne patches with
$\tdu \tdv >0$.
Each instanton phase of the cycle has  $\epsilon=1$  and therefore
$k=-1$. It lifts to   
 \be
 ds^2_{d+1}= \left(1 - {\tdu _0^2 \over  r ^{d-2}}\right) dy^2- \left(1 - {\tdu _0^2 \over  r^{d-2}}\right)^{-1} dr^2 
   + r^2dH^2_{d-1} \, .
\label{cvbb}
\ee
This is equivalent to  (\ref{cvb}) with $m=\tdu _0^2$.

Let us now follow a cycle  as $\lambda $ is varied from $0$ to $2\pi $. 
At $\lambda =0$ ($\tdu =0$) we have a big bang singularity, which lifts to the
curvature singularity at $r=0$ of the $d+1$ black hole.
The universe expands and then collapses to a big crunch singularity 
at $\lambda =\pi/2$ ($\tdv =0$), which lifts to the horizon of the $d+1$ black
hole at $r^{d-2}=\tdu _0^2$. For  $\lambda \in ({\pi\over 2},\pi )$ the solution  lifts
to (\ref{cvbb}). Note, however, that the $(d+1)$-metric changes {\it discontinuously}
across the horizon; this can be seen from the fact that the near horizon geometry of the black hole 
differs from the near horizon geometry of the solution (\ref{cvbb})
because the $(d-1)$-sphere of the former becomes a $(d-1)$-hyperboloid
of 
the latter. 
Thus, in contrast to the $\gamma=1 $ case, the trajectories that are smooth for 
$\gamma=2$ do not lift to solutions in the higher dimension that 
are smooth across a horizon. This makes the higher-dimension interpretation
of cyclic cosmologies problematic, and indicates a `higher-dimensional preference' for
the Milne geodesics that we found for $\gamma=1$. 

If we put this difficulty  aside and continue with the cycle, then we approach the curvature singularity (at $r=0$) of the 
solution (\ref{cvbb}) as  $\lambda$ increases to  $\pi $.  Passing through
$\lambda=\pi$, we arrive at the second cosmological phase of the cycle,
with  $\lambda \in ({\pi},3\pi/2 )$, described again in $d+1$ dimensions by the 
interior of the black hole. Passing again through a big crunch singularity at $\lambda= 3\pi/2$ 
we come to the second instanton phase, in the interval  $\lambda \in ({3\pi\over 2},2\pi )$, which  
again lifts to the $d+1$ metric (\ref{cvbb}).

\section{Uplift to (anti) de Sitter Schwarzschild}
\setcounter{equation}{0}

Although cosmological trajectories of the $\Lambda\neq 0$ models 
considered in this paper can be understood  qualitatively for any value of the constant $a$, there is only one other value for which exact solutions can be readily found (leaving aside the $a=0$ case, which we discuss in the following section). 
This is because there is only one value of $a$ for which the 
equations of motion
for $U$ and $V$ separate. From (\ref{masteract}) it can be seen that
the 
potential terms depend only on $U$ iff
\be
\gamma=1\, ,\qquad  2 \alpha^2 \gamma -1 = \alpha \gamma a\, , 
\ee
which together require that
\be
 a = \sqrt{\frac{2}{(d-1)(d-2)}} \,. 
 \ee
The effective action (\ref{masteract}) in this case is
\be
I= -{1\over2}(d-1) \int d\tau \left\{ \dot    U
\dot   V -4k  + {4\Lambda \over (d-1)(d-2)} \left(
 U ^2\right)^{1/(d-2)}\right\}\, . 
\ee
The field equations are now
 \be
  \ddot  { U} = 0 \,, \quad  \ddot  { V} =  \frac{8
  \Lambda ( {U}^2)^{1/(d-2)}}{(d-1)(d-2)^2  U} \,,
 \ee
and the zero-energy constraint is
 \be
 \dot  U \dot   V =- 4 k + \frac{4  \Lambda ( {U}^2)^{1/(d-2)}}{(d-1)(d-2)} \,.
 \ee
The most general solution is the curve
 \be
   U = c \tau  \,, \quad
   V = -\frac{4 k \tau}{c} + m + \frac{4 
  \Lambda ((c \tau)^2)^{1/(d-2)} \tau}{c d (d-1)} \,, 
 \ee
 for constants $c,m$, as in the special case of $\Lambda=0$ that we
 considered in the previous section. Following the discussion of the uplift to 
 $(d+1)$ dimensions in that section, we set $c\tau = r^{d-2}$ 
 (in regions with $U>0$) to again get the metric 
 \be\label{Schw2}
  ds_{d+1}^2 =  h(r)^{-1} d r^2 - h(r) dz^2 + r^2 d \Sigma_k{}^2 \,,
 \ee
 but now with a different function $h$. Choosing $c=2$, one finds that
  \be
 h( r) = k -{m\over r^{d-2} } - {\Lambda\over d(d-1 ) }\, r^2 \, . 
\ee
For $k=1$, this is the (anti-)de Sitter Schwarzschild metric. 

A new feature with respect to the $\Lambda=0$ case is that the
de Sitter-Schwarzschild metric that arises for $\Lambda>0$ has
two horizons for sufficiently small non-zero $\Lambda$. 
When this happens, the region between the horizons is the
lift of an instanton solution which mediates a universe collapsing to
a big crunch singularity and one expanding from a second big bang singularity.
To illustrate this, we consider an example with $d=4$ and $k=1$.
In this case the $d+1$ solution is the five-dimensional de Sitter
Schwarzschild black hole with 
\be
  h( r) = -{\Lambda \over 12 r^2} (r^2-r_+^2)(r^2-r^2_-)\ ,
\ee
where
\be
r_\pm^2 ={6\over \Lambda }\left( 1\pm \sqrt{1-{m\Lambda\over 3} }\right)\, . 
\ee
There are two horizons for $m\Lambda< 3 $. The horizon at $r=r_-$ is the black hole horizon, while 
$r=r_+$ is the de Sitter cosmological event horizon.

Let us consider the cosmological evolution for the cases:
a) $m=0$; b) $0<m\Lambda<3 $; c) $m\Lambda>3 $ (see fig.~1).

\begin{figure}[ht]
\centerline{
\epsfig{file=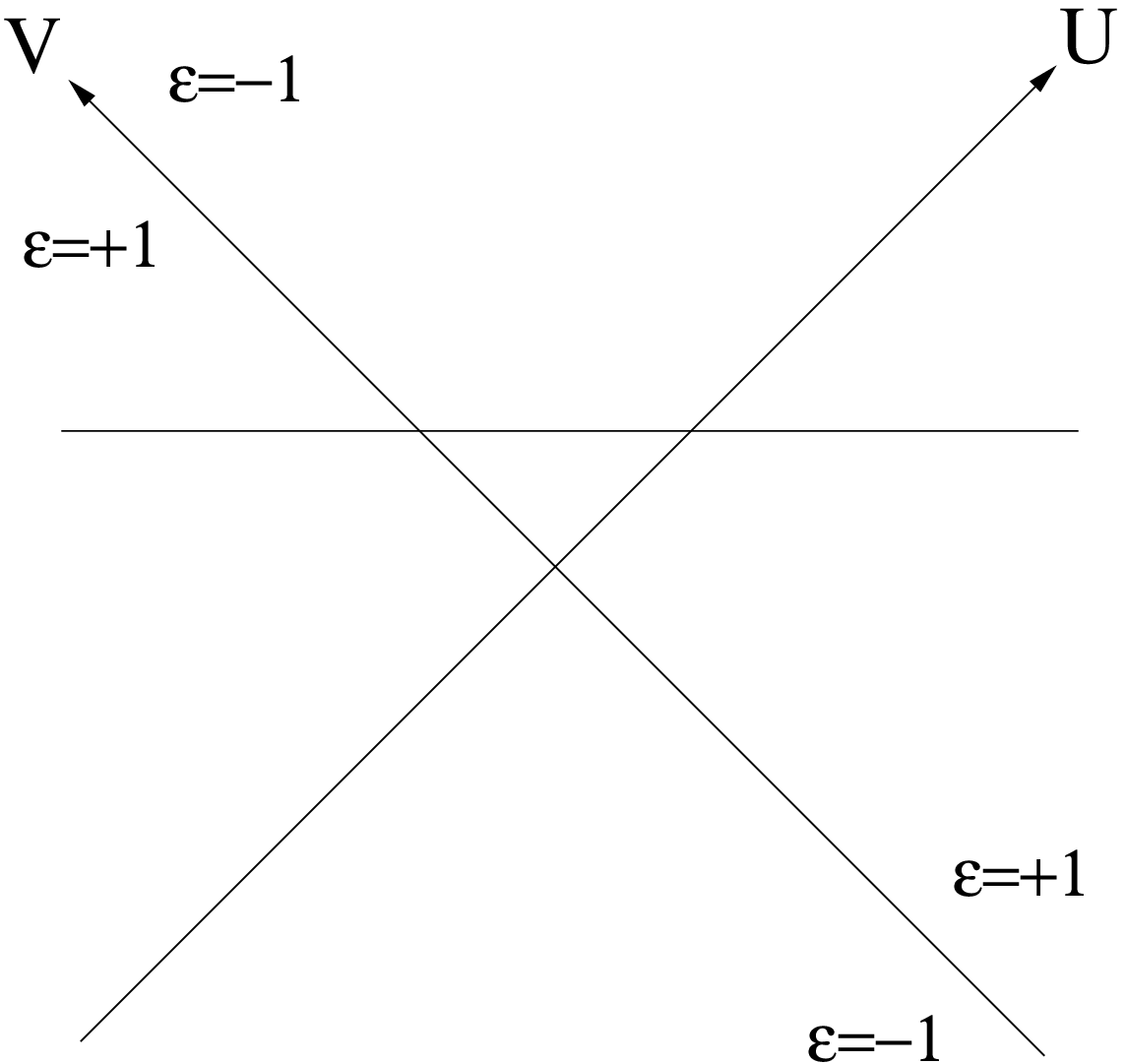,width=.36\textwidth}
1a
\qquad\qquad
\epsfig{file=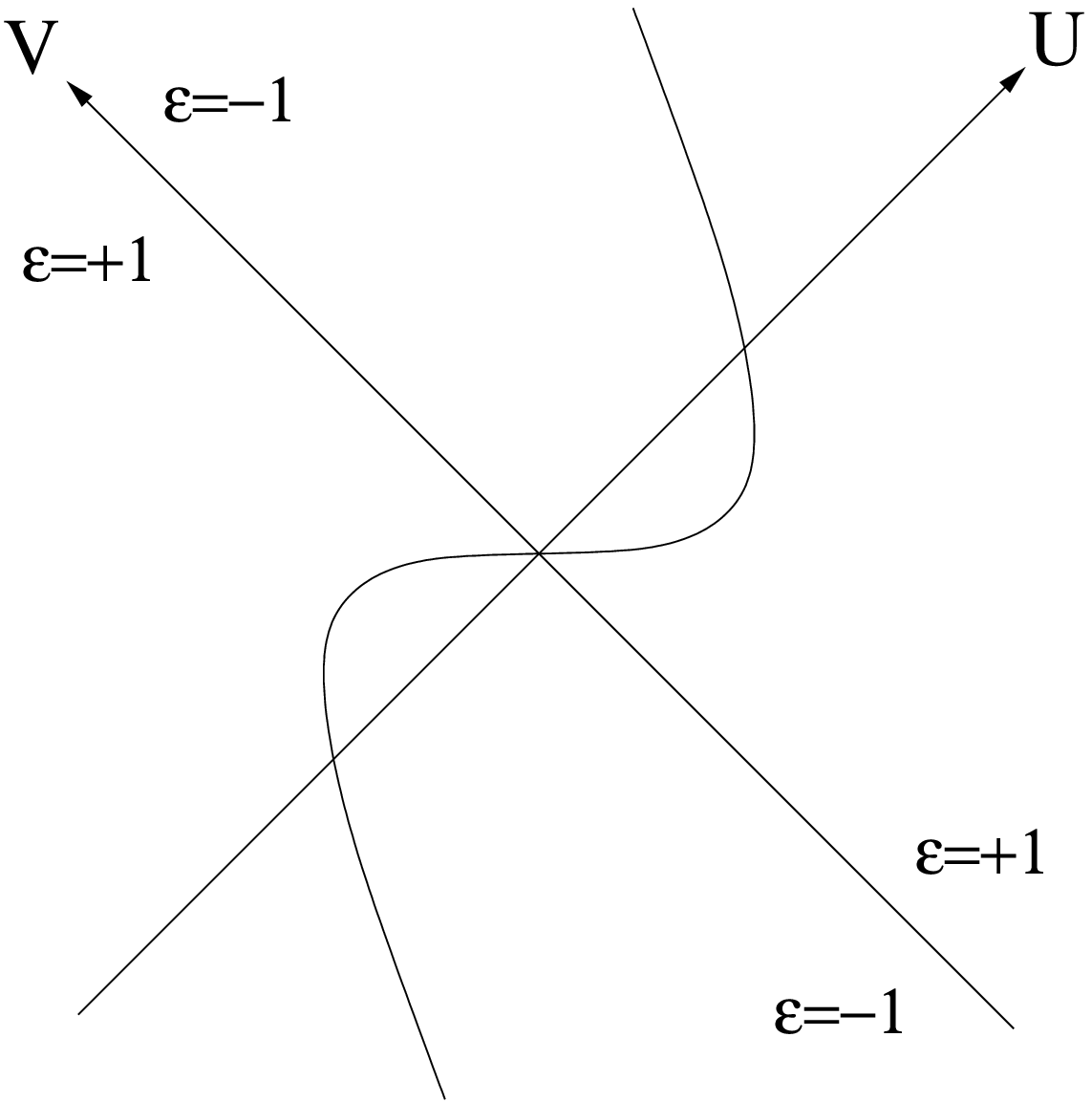,width=.36\textwidth}
1b }

\medskip\medskip
\centerline{
\epsfig{file=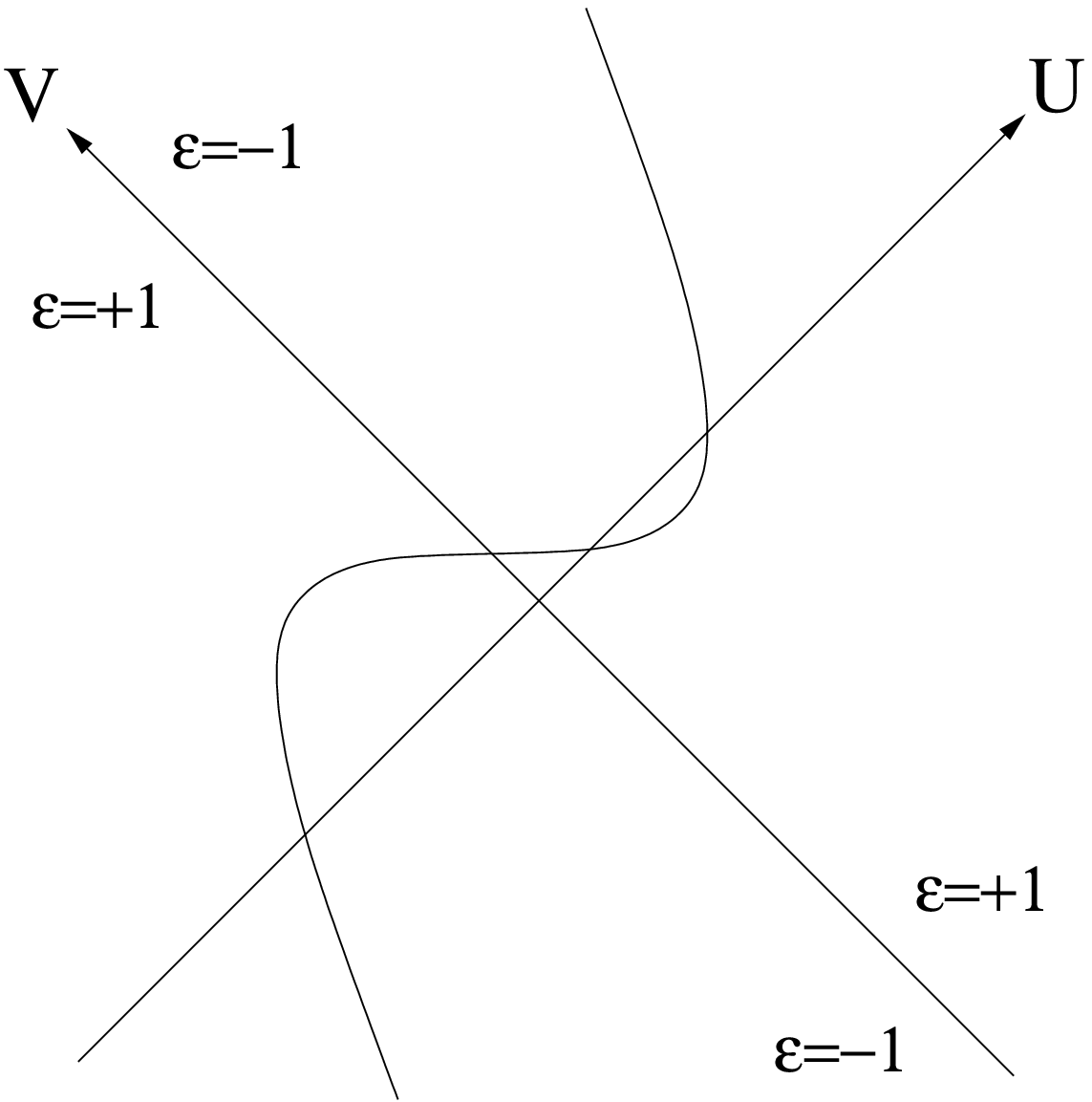,width=.38\textwidth}
1c\qquad\qquad
\epsfig{file=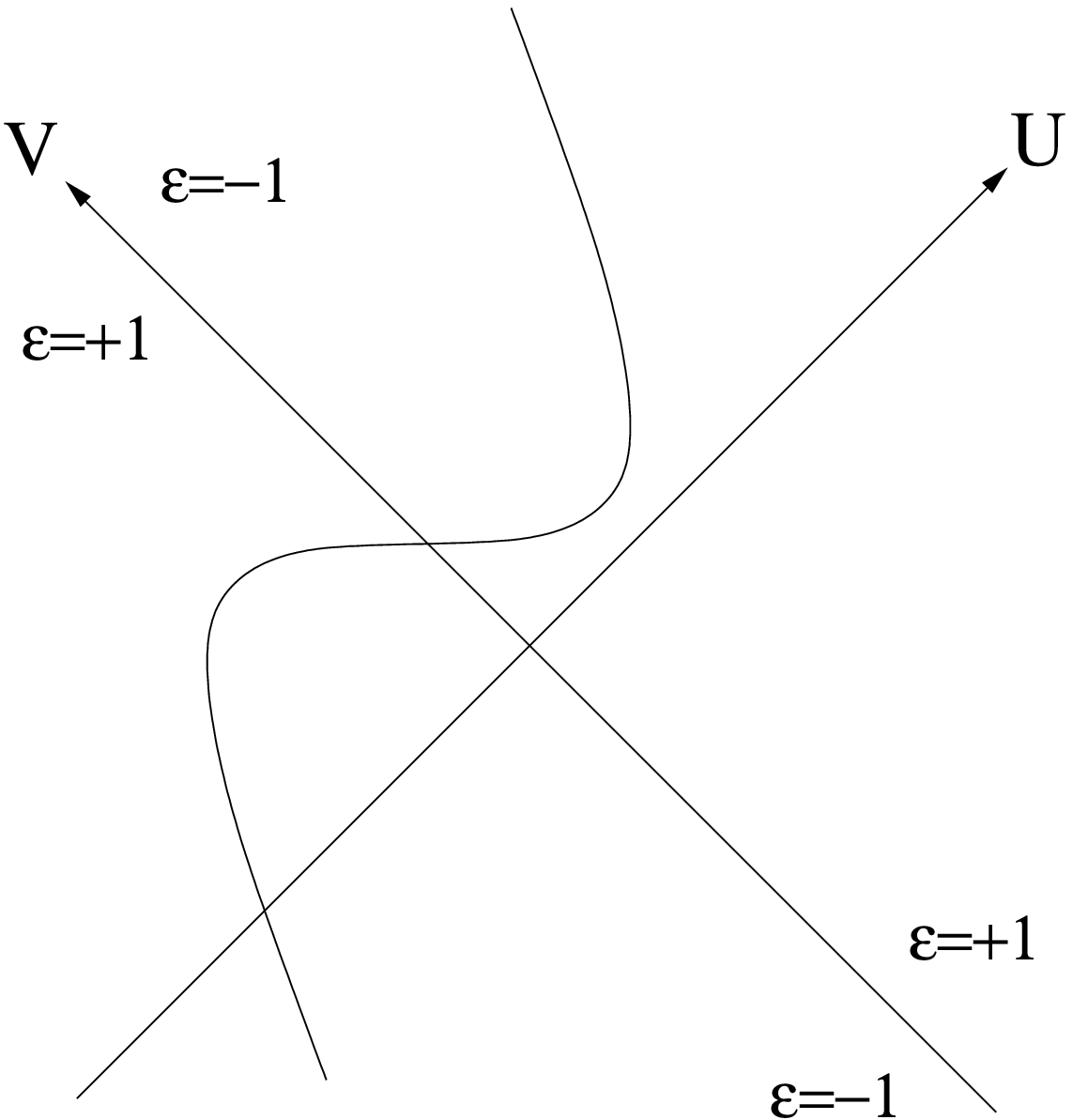,width=.39\textwidth}
1d}
 \caption{\it Generic trajectories for $k=1$ and $m \geq  0$: a) $\Lambda=0$, b)
   $\Lambda>0$ and $m=0$, c) $0 < m \Lambda <3$, d)  $m \Lambda > 3$. The behaviour near
$ U=0$ is determined by $k$ and $m$ while the asymptotics depend on
$\Lambda$.}
\end{figure}

\begin{itemize}

\item $\Lambda=0 $. This case was analysed in the previous section. 
The trajectory is just a straight line.

\item $m=0$. In this case the trajectory 
passes through the origin $ U= V=0$ and is symmetric
under $ (U,V) \to - (U,V)$. Starting at negative $ U$ and $ V$, we may interpret 
the trajectory as representing a contracting universe that collapses 
through a big-crunch singularity into an instanton phase,
which is connected to an isometric instanton phase through the
origin $U= V=0$. Then, as the trajectory passes again through the
Milne horizon at $ V=0$ it becomes an expanding big-bang universe,
this being the time reverse of the original big-crunch universe.

\item $0<m\Lambda<3 $. Starting from the 
big-bang singularity at $\tau =0$ ($ U=0$),
the universe
expands  and then collapses to a big crunch singularity at
$2\tau=r_-^2$ ($  V=0$). This evolution takes place in the future
Milne sector ($  U>0, \,   V>0$). For $2\tau \in (r_-^2,r_+^2)$, the
trajectory is in the Rindler region with $  U>0$,\ $  V<0$,
and the corresponding solution is therefore an instanton.
This instanton lifts to the $d+1$ spacetime between the black hole and
de Sitter horizons.
As $2\tau $ increases to $r_+^2$, the trajectory returns to the
$  V=0$ Milne horizon.
Having passed through this horizon, the trajectory is once more in 
the future Milne sector  and represents a second
expanding big-bang cosmology. This universe expands forever
and  at late times  approaches the self-similar universe obtained by dimensional reduction of
the five-dimensional de Sitter universe.

The full trajectory, which is shown in fig. 1c, includes another big-crunch to big-bang transition 
through an instanton that lifts to the interior spacetime of the negative mass Schwarzschild de Sitter spacetime. The exterior of this spacetime is the lift of a universe that is collapsing to a big crunch singularity, which lifts to the cosmological horizon of the negative mass de Sitter-Schwarzschild  spacetime.

\item
$m\Lambda =3$.
In this case the black hole solution becomes the so-called Nariai solution
with the topology $dS_2 \times S^{d-1}$.
In the future Milne patch we get a
bouncing cosmology where the singular bouncing point gets resolved into the
$dS_2$ part of the Nariai solution.

\item $m\Lambda>3 $. The part of the trajectory with $  U<0 $ is similar to the previous case.    
Starting from the big-bang singularity at $\tau =0$ ($ U=0$), the universe first
expands but then collapses to a minimum size, after which it re-expands, beginning
a phase of eternal  expansion.

\end{itemize}

\medskip

The cases with $k = -1$ are related to those
with $k = 1$ by flipping the sign of $\Lambda, \ m $ and $V$: 
the evolution is similar with instantons and cosmologies interchanged
(since $V$ flips sign).

\section{Cosmological Constant models ($a=0$)}
\setcounter{equation}{0}

We now allow for non-zero $\Lambda$ but set $a=0$. In other words, we now consider  a
cosmological constant term, of either sign. The effective action (\ref{masteract}) becomes
 \be
 \label{zeroa}
  I =- {1\over2}(d-1)\int d\tau \,\Big\{ \gamma^2 \dot U \dot V -  4 k (-\epsilon U V)^{\gamma - 1}
  + \frac{4  \Lambda}{(d-1)(d-2)} (-\epsilon U V)^{2 \alpha^2 \gamma  -1} \Big\}\, . 
 \ee
We consider first the case of general $d$, showing that our prescription for passing 
through cosmological singularities leads fairly generically to cyclic universes. 
We then show how exact solutions may be found for $d=3$ and $d=4$. 

\subsection{Cyclic universes} 

For $a=0$ we have a problem analogous to that of a particle in a central potential. 
To make this explicit, we set
\be
U= \eta e^{-\psi}\, , \qquad V= -\epsilon \eta e^\psi\ ,\qquad U>0\ , \nonumber
\ee
\be
U= -\eta e^{-\psi}\, , \qquad V= \epsilon \eta e^\psi\ ,\qquad U<0\ ,
\ee
to put the action into the form
 \be
 \label{zerob}
  I = -{1\over2}(d-1)\epsilon \int d\tau \,\Big\{ \gamma^2
  \left(\dot \eta^2  - \eta^2\dot\psi^2\right) 
+ 4k\epsilon \,\eta^{2(\gamma-1)} - 
\frac{4 \epsilon \Lambda}{(d-1)(d-2)} \eta^{4 \alpha^2 \gamma  -2} \Big\}\, . 
 \ee
 Let us note here that the relation between the new variables  $(\eta,\psi)$ and the original 
 variables $(\varphi,\phi)$ is
 \be
 e^\varphi = \eta^{2\alpha\gamma}\, ,\qquad \phi = 2\alpha\gamma \psi\, , 
 \ee
 and that the spacetime metric in terms of $\eta$ is
 \be\label{spacetimemetric}
 ds^2_d = {4\epsilon\over (d-2)^2} \left(\eta^2\right)^{2\alpha^2\gamma-2} d \eta^2
 + \left(\eta^2\right)^{2\alpha^2\gamma/(d-1)}d\Sigma_k^2\, .
 \ee
 
 The potential term is greatly simplified by the choice
$\gamma=1$. Omitting an unimportant overall factor\footnote{
This includes a factor of $\epsilon $, which ensures that there is no
factor of $\epsilon $ in the kinetic term in the $U,\, V$ variables
that cover the full 2d Minkowski space.}, this choice yields the Lagrangian 
\be
L= \dot  \eta ^2 - \eta^2\dot \psi^2 + 4k\epsilon
-{4\epsilon\Lambda \over (d-1)(d-2)}
\eta^{2/(d-2)}\, .
\ee
The $\psi$ equation of motion is trivially once-integrated to give
\be
\dot  \psi = {j\over \eta^2}\, , 
\ee
for some constant $j$. The $\eta$ equation is
\be
\eta \ddot \eta =- {j^2\over \eta^2} - {4\epsilon\Lambda \over
  (d-1)(d-2)^2}\eta^{2/(d-2)}\, . 
\label{jnn}
\ee
This is trivially once-integrated and the integration constant is fixed by the 
zero-energy constraint. This yields the equation\footnote{This equation implies
the equation of motion (\ref{jnn}) except when $\dot \eta
\equiv 0$, in which  case both the constraint and the $\eta $ 
equation  of motion are needed. }
\be
\dot \eta^2 +V_{\rm eff}(\eta )=4k\epsilon\, ,
\label{adf}
\ee
where
\be
V_{\rm eff}(\eta )=  - {j^2\over \eta^2} + {4\epsilon\Lambda \over (d-1)(d-2)}\eta^{2/(d-2)}\, . 
\label{adff}
\ee
The cosmological  ($\epsilon=-1$) solutions to this system  are well
known for $j=0$; the dilaton is constant, and the metric is de Sitter or 
anti de Sitter space. In what follows we assume $j\neq 0$.

Our main interest at present is the possible continuation through a
cosmological singularity to an instanton ($\epsilon=1$) solution. 
As cosmological singularities occur at $\eta =0$, the term containing
the cosmological constant in eq.~(\ref{adf}) is irrelevant near the
singularities. It follows that near a singularity the trajectories
are straight lines in the $(U,V)$ plane, so the transition through the
singularity is exactly the same as in the $\Lambda =0$ case already
discussed. Away from the singularity, the trajectories are bent by the
cosmological constant term and the global behaviour can be quite
different from the $\Lambda =0$ case. Nevertheless, it can be understood 
by examining the behaviour of a particle in the
potential $V_{\rm eff}(\eta )$, noting that the potential changes across
the Milne horizons because of the change in the sign of $\epsilon $.

Consider first $\Lambda>0$.
{}For an instanton, the potential is monotonically increasing with
increasing $\eta $, going to $-\infty$ at $\eta =0$ and $+\infty $ at $\eta=\infty$.
So any trajectory will flow to $\eta=0$ and cross the Milne horizon,
to yield a big bang cosmology.
What happens next depends on $k $ and $\Lambda $.
On the cosmology side of the Milne horizon the potential is
everywhere negative and has a maximum. 
If this maximum is less than the energy $4k\epsilon$
then  $\dot \eta $ is never zero and the universe expands forever.
This happens when  $k=-1$ for any $\Lambda>0$ and also for $k=1$ provided
that
\be
\Lambda > \Lambda _{+}  =(d-2)^2 \left({4\over (d-1)j^2}\right)^{1/(d-2)}\, . 
\ee
If, instead, the maximum is greater than the energy $4k\epsilon$,
which happens when $k=1$ and $\Lambda < \Lambda _+$, then there
will be a recollapse to a big crunch singularity. However, the
trajectory will pass straight through the singularity to the region
behind the Milne horizon and we will have a second instanton phase.
This instanton trajectory will be bent back to the Milne horizon at
$\eta=0$, yielding a second cosmological phase identical to the first
one. Thus we have a cyclic universe, as is illustrated in figure~2.

\begin{figure}[ht]
\centerline{
\epsfig{file=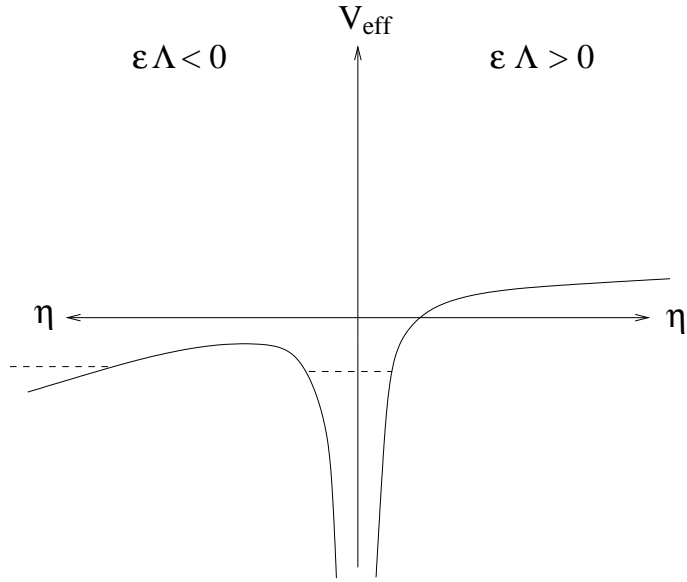,width=.60\textwidth}}
 \caption{\it The potential $V_{eff} (\eta)$ with $j \neq 0$ for both signs
of $\epsilon \Lambda$. The variable $\eta$ is positive on both sides. A
cyclic universe corresponds to a particle in the potential well around $\eta =0$ (with dashed energy level).
Classically it is trapped in the well; quantum-mechanically it could tunnel through the energy barrier to the left region.}
\end{figure}

A similar analysis applies for $\Lambda <0$. If $k=1$ or if $k=-1$ 
with $\Lambda < \Lambda _{-} = -\Lambda_{+}$, then we have an 
instanton phase that yields a big bang cosmology, which recollapses through 
a big crunch to another  instanton phase. If $k=-1$ but $\Lambda > \Lambda _{-} $, 
then this process continues {\it ad infinitum} and we have a cyclic universe.
Thus cyclic universes are generic in this model.

\subsection{Explicit examples} 
The constraint (\ref{adf}) can be written as 
\be
d\tau = {d\eta\over \sqrt{4k\epsilon -V_{\rm eff}(\eta)}}\, . 
\ee
Using this in (\ref{spacetimemetric}) we have\footnote{
Related solutions have appeared in the literature
(see e.g. \cite{myers}).
Here our main interest is the fact that there exist smooth
transitions through the
Milne horizon connecting instanton and cosmology solutions.}
\be
ds^2_d= {4\epsilon \over (d-2)^2}\, {\eta^{-{2(d-3)\over d-2}}\over
  \left[4k\epsilon-V_{\rm eff}(\eta )\right]} d\eta^2 +\eta^{2/(d-2)}d\Sigma_k^2\ ,
\ee
so that $\eta$ is now the time variable. The dilaton  is given 
as a function of $\eta$ by the formula
\be
\phi ( \eta )=2\alpha \, j\int^\eta {dx\over x^2\sqrt{4k\epsilon-V_{\rm eff}(x 
    )}}\ .
\ee

More explicit formulae with $\tau$ as the time variable 
can be obtained for $d=3$ and $d=4$. 
For $d=3$ the equation (\ref{adf}) becomes
\be
\dot  \eta^2 = 4k\epsilon + {j^2\over \eta^2}
-2\epsilon\Lambda \eta^2\ .
\ee
Setting 
\be
y=\eta^2-\frac{k}{\Lambda} \qquad {\rm and} \quad b =
\frac{k^2}{\Lambda^2}+\frac{j^2}{2 \epsilon \Lambda}\, ,
\ee
this equation becomes
\be
\dot  y^2 = -8\epsilon\Lambda \left(y^2-b\right)\,.
\ee
This can be easily solved, and the solutions for the different cases
are as follows:
\begin{itemize}
 \item $\Lambda \epsilon <0$
\begin{align}
\eta^2 &= \sqrt{b} \cosh\left(\sqrt{-8 \Lambda \epsilon} (\tau-\tau_0)
\right)+\frac{k}{\Lambda} &{\rm for} \quad b>0\, , \nonumber\\
\eta^2 &= \sqrt{-b} \sinh\left(\sqrt{-8 \Lambda \epsilon}( \tau-\tau_0)\right)+\frac{k}{\Lambda} &{\rm for} \quad b<0\,,
\end{align}
where $\tau_0$ is an integration constant. For $b>0$ with $k/\Lambda>0$, the domain of validity for $\tau$ is $(-\infty, \infty)$. For the rest of the cases, $\tau_0$ be chosen such that the domain of validity is $(0, \infty)$.
\item $\Lambda \epsilon >0$
\begin{equation}
\eta^2 = \sqrt{b} \sin\left(\sqrt{8 \Lambda \epsilon}(\tau-\tau_0) \right)+\frac{k}{\Lambda}\,.
\end{equation}
Since, by definition, $\sqrt{b}>k/\Lambda$ for $j \neq 0$, there is always a non-zero domain of validity for this solution.
\end{itemize}

Note that in this case of  $d=3$ the time $\tau $ coincides with the  
FLRW   cosmic time  $t$, as can be seen from  eq.(\ref{FLRWtime2})
using  $\alpha=1 $ and $\gamma=1$, and $\eta (\tau )$ is the FLRW scale factor.

\medskip

In the case of $d=4$, eq.~(\ref{jnn}) becomes
\be
\eta \, \dot \eta = \sqrt{   {j^2}+4k\epsilon\, \eta^2  -
  {2\over 3}\epsilon\Lambda \, \eta^{3}} \, . 
\label{poi}
\ee
The solution  $\eta (\tau ) $ can be expressed in terms of elliptic functions but the explicit formula is not illuminating and we omit it.

\section{Flat  universes ($k=0$)}
\setcounter{equation}{0}

In this section we consider flat universes
($k=0$) for which the solutions are known exactly for any value of $a$
\cite{Burd:1988ss,Townsend:2003qv,Russo:2004ym}. Here we present these
solutions as straight lines in a 2D Minkowski space, and consider
the continuation through the cosmological singularities at the Milne
horizon.

Going back to our original scalars $\phi$ and $\varphi$, the $k=0$ effective action reads
 \be
I=  \int d\tau\,\left\{
 \tfrac{1}{2} f^{-1} \epsilon \left( \dot\varphi^2 - \dot\phi^2
 \right) - 
f \Lambda e^{2\alpha\varphi - a\phi} \right\}\,. \label{actionkzero}
 \ee
It is convenient to introduce the quantity
\be
\Delta = a^2 - 4\alpha^2\, , 
\ee
because this is invariant under $SO(1,1)$ rotations of
 $(\varphi,\phi)$. We will assume initially that $\Delta\ne0$, in
 which case 
we may define  $s$ by  $s = {\rm sign} ( \Delta)$. 
We now define new variables $(\zeta,\xi)$ by means of an $SO(1,1)$ rotation:
\begin{equation}
\begin{pmatrix}
{\zeta} \cr
{\xi}
\end{pmatrix} = {1\over \sqrt {|\Delta| }}
\begin{pmatrix}
2\alpha & -a\cr
- a & 2\alpha\cr
\end{pmatrix}
\begin{pmatrix}
\varphi\cr
\phi
\end{pmatrix} \ \ .
\label{mtx}
\end{equation}
This brings the action (\ref{actionkzero}) to the form
 \be
 I= \int d\tau\,\left\{
 \tfrac{1}{2} f^{-1} s \epsilon \left(\dot {\zeta}^2 - \dot{\xi}^2 \right)
 - f \Lambda e^{\sqrt{|\Delta |} {\zeta} } \right\}\,. \label{ooo}
 \ee
By analogy with  (\ref{kop}) we choose new coordinates $(\U,\V)$ 
 \be
 e^{\zeta} = (- \epsilon\, \U \V )^{\alpha \gamma} \,, \qquad
 e^{ \xi} = \left( -\frac{ \epsilon \V }{\U } \right)^{\alpha \gamma} \,, 
\label{gty} 
\ee
and we fix the time-reparametrization invariance by the choice
\be
 f = {2 \alpha^2 \gamma^2 \over  s \epsilon\, \U \V }\, .
\ee
The action becomes
 \be\label{masteract2}
  I = \int d\tau\,\Big\{ \dot \U \dot \V
  + 2 \alpha^2 \gamma^2  s \Lambda  (- \epsilon \, \U \V )^{\alpha \gamma\sqrt{|
      \Delta | }     -1} \Big\}\, ,
\ee
and the $f$-equation of motion of (\ref{ooo}) becomes the constraint
 \be
 \dot \U \dot \V  =  2 \alpha^2 \gamma^2  s\Lambda  
 (- \epsilon \, \U \V )^{\alpha \gamma\sqrt{| \Delta|}  -1}\, .
 \ee
With the choice $\gamma = 1/(\alpha \sqrt{| \Delta |})$,
 the solutions to the field equations satisfying the constraint are
 the straight lines\footnote{
Note that the choice
  $\gamma = 2/(\alpha \sqrt{| \Delta |})$ also gives rise to linear field equations.
 In this case, for $s\epsilon \Lambda>0$, the  solution to the field equations will be a segment of an ellipse,
but a smooth continuation through the Milne horizon would require
changing the sign of  $s \Lambda$, which is not possible because
$s$ and $\Lambda $ are parameters of the model.}
\be
\U =c\tau\ ,\qquad \V ={2 s\Lambda\over c| \Delta|}\, \tau+m\ .
\ee
In order to find the general metric and dilaton solution explicitly,
we need to express $U,\ V$ appearing in eq.~(\ref{einstein})
in terms of $\U,\ \V$. Using eqs.~(\ref{kop}), (\ref{mtx}) and  (\ref{gty}), we
find
\be 
\U^2=\left( U^2\right)^{2\alpha+a\over \sqrt{| \Delta|}}\ ,\qquad
\V^2=\left( V^2\right)^{2\alpha-a\over \sqrt{| \Delta|}}\ .
\ee

The analysis of the continuation through the Milne horizon in the
$(\U,\V )$ space is similar to the discussion  of the $\Lambda=0$
case in section 3. The role of $k$ in that section is  played here by
$- \alpha^2 s \Lambda/2$
(which, however, need not be $\pm 1$).
Consequently, the full straight
line trajectory consists of cosmological and instanton phases.

\medskip

The case $\Delta = 0$, i.e.~$a = 2 \alpha$, must be treated
separately. 
We choose $\gamma=1/\alpha^2 $ and
we define new coordinates $(U,v)$ by  
 \be
  e^\varphi = (\epsilon U)^{1/\alpha } e^{ v} \,, \qquad
  e^\phi = (\epsilon U)^{-1/\alpha } e^{ v} \,.
 \ee
In the $(U,v)$-plane, the regions with $U<0$ represent cosmologies,
while the  regions with $U>0$ represent instantons.

The gauge choice
 \be
  f = \frac{2 }{\alpha \epsilon U} \,,
 \ee
yields the following action
 \be
  I = \epsilon 
\int d\tau\,\Big\{ \dot U \dot v - {2\Lambda  \over \alpha}  \Big\} \,,
 \ee
and the additional constraint
 \be
  \dot U \dot v = - {2\Lambda  \over \alpha}   \,.
 \ee
Again we have straight lines in the $(U,v)$-plane, but now
 cosmological singularities occur at points at which the trajectory
 crosses
$U=0$. As every trajectory crosses $U=0$ once,
every trajectory has one instanton and one cosmology phase.

\section{Discussion} \setcounter{equation}{0}

In this paper we have investigated the possibility of a classical resolution 
of  cosmological singularities in a class of $d$-dimensional models 
with a dilaton field $\phi$ and a possible exponential potential of the form 
$\Lambda e^{-a\phi }$. In these models a FLRW cosmology corresponds to a 
trajectory in a 2D Minkowski `superspace'  but with a Milne metric 
in  the parametrization provided by the 
dilaton $\phi$ and the cosmological scale 
factor $e^\varphi$. In terms of null Minkowski coordinates $(U,V)$ these fields
cover only the Milne patches, for which $UV>0$. Cosmological singularities 
correspond to points on trajectories that cross the Milne horizon at $UV=0$.
Although the $d$-dimensional spacetime metric is singular, the cosmological 
trajectory can cross the Milne horizon smoothly, and the continuation of the trajectory into 
the Rindler region behind the Milne horizon represents an instanton solution of 
the Euclidean action. Thus, a trajectory that starts in a Milne region, smoothly 
crosses to  a Rindler region and then returns to a Milne region
represents a big crunch-big bang transition mediated by an instanton phase.
Globally, the trajectories may be open curves that represent a solution with a
finite number of cosmology/instanton phases, or closed curves 
that represent cyclic 
universes.

These results  complement our earlier study of cosmology/instanton transitions in 
models of gravity coupled to scalar fields with a hyperbolic target
space \cite{Bergshoeff:2005cp}.  Indeed, for $\Lambda=0$, the model we considered is a
consistent truncation of the model of  \cite{Bergshoeff:2005cp}  
to purely Einstein-dilaton system.
However,
an important aspect of the truncated theory considered here is that its action is the 
dimensional reduction of the Einstein-Hilbert action in one higher dimension. This is
 true not only for the standard metric/dilaton action, for a Lorentzian-signature metric, but also for the Euclidean action, which is obtained by 
considering a time-independent metric in the higher dimension. Thus, the cosmology/instanton solutions lift to solutions of Einstein's equations in one higher dimension. For $k=1$ a cosmology/instanton solution corresponding to a straight line trajectory in the 2D Minkowski space passes through both Rindler patches and either the future or past Milne patch of this space; the full trajectory therefore comprises three segments. The cosmological solution corresponding to the segment in the Milne patch  lifts to the interior of a (positive mass) Schwarzschild black hole, while the  pre-big-bang and  post-big-crunch instanton solutions corresponding to the segments in the Rindler patches lift either to the exterior of the same black hole solution or to the negative mass Schwarzschild  spacetime found by reversing the sign of the black hole mass. All three of these higher-dimensional spacetimes (for given mass)  become part of the {\it same} cosmology/instanton solution in one dimension lower.
In particular, the black hole horizon becomes either the big bang or the big crunch singularity
of the lower-dimensional cosmology, depending on which of two possible
dimensional reduction ans\"atze  is used, the two ans\"atze being
related by a change of sign of the dilaton\footnote{
A interesting corollary  is a means of resolving the curvature
 singularity of the Schwarzschild solution: reduction by one dimension
 followed by a change of sign of the dilaton and a lift back to the
 higher dimension interchanges the positive and negative mass
 Schwarzschild solutions and also interchanges the black hole horizon
 with the curvature singularity behind the horizon.}.

Motivated in part by massive IIA supergravity, for which the dilaton has a positive exponential 
potential, we also considered the effect of an exponential dilaton potential parametrized by a 
constant $a$, such that $a=0$ corresponds to a constant potential (i.e. a cosmological constant).
For given $d$ there is one value of $a$ for which exact cosmology/instanton solutions can be found for any $k$;  namely the value for which the action is the reduction of gravity with a cosmological constant in one higher dimension. These solutions lift to de Sitter-Schwarzschild (for positive cosmological 
constant) or anti-de Sitter-Schwarzschild (for negative cosmological
constant). The de Sitter Schwarzschild case is particularly
interesting because of the occurence (for sufficiently small positive
black hole mass) of two horizons, the black hole horizon and the cosmological event horizon. 
These horizons are the lift, respectively,  of a big crunch and big bang singularity on a 
trajectory that crosses and then recrosses the Milne horizon of the 2D Minkowski `superspace'.
The region behind the Milne horizon is described by a $d$-dimensional 
instanton solution mediating the big crunch to big bang transition, and this lifts to the 
$(d+1)$-dimensional spacetime between the black hole and cosmological horizons. 
Thus, {the entire big crunch to big-bang transition via a cosmological
  instanton lifts to a non-singular spacetime in one higher
  dimension}!  
The uniqueness of the analytic continuation of the $d+1$ dimensional spacetime
through its horizons is evidence for the correctness of
our prescription for the continuation of cosmological trajectories
through their singularities.
Moreover, since the horizons remain regular when small perturbations
about spherical symmetry are included, this suggests that it may be
possible to extend our prescription for continuation through
cosmological singularities
to more general cosmologies which are not homogeneous and isotropic.

\section*{Acknowledgements}

E.B. would like to thank Jelle Hartong for a useful discussion.
E.B. and A.C.  and J.R. acknowledge partial  support
by the  European Commission FP6 program
MRTN-CT-2004-005104. J.R. also acknowledges support from
MCYT FPA 2004-04582-C02-01 and CIRIT GC 2001SGR-00065. 
The research of D.R. is funded by the PPARC grant PPA/G/O/2002/00475.

\setcounter{section}{0}
\bigskip



\end{document}